Competing Atomic and Molecular Mechanisms of Thermal Oxidation

Xiao Shen[1], Blair Tuttle[1], and Sokrates T. Pantelides[1,2,3]

[1]Department of Physics and Astronomy, Vanderbilt University, Nashville, TN 37235

[2]Department of Electrical Engineering and Computer Science, Vanderbilt University, Nashville, TN 37235

[3]Materials Science and Technology Division, Oak Ridge National Laboratory, Oak Ridge, TN 37831

**Abstract**

The oxidation of SiC and Si provide a unique opportunity for studying oxidation mechanisms because the product is the same, $SiO_2$. Silicon oxidation follows a linear-parabolic law, with molecular oxygen identified as the oxidant. SiC oxidation obeys the same linear-parabolic law but has different rates and activation energies and exhibits much stronger face-dependence. Using results from first-principles calculations, we show that atomic and molecular oxygen are the oxidant for Si- and C-face SiC respectively. Comparing SiC with Si, we elucidate how the interface controls the competition between atomic and molecular mechanisms.

Oxidation is a ubiquitous process that occurs in most solids including metals, insulators, and semiconductors. The atomic-scale processes that underlie oxidation are of fundamental interest as they control the quality of both the resulting oxide and its interface with the substrate. Furthermore, elucidation of these processes would also benefit applications as thermal oxidation is widely used to fabricate oxide films, such as gate dielectrics and insulating layers, in electronic devices, [1,2] in nanostructures, [3] and in applications of ceramics materials [4]. The best-known example is Si oxidation, which produces a high-quality oxide/semiconductor interface and is one of the most important techniques that enable semiconductor technology. Meanwhile, for materials used at high temperatures, stability against thermal oxidation is a major concern [5].

Silicon and SiC provide a unique opportunity for investigating thermal oxidation because they have the same native oxide, $SiO_2$. Si oxidation has been extensively studied and the kinetics has been well described by a model proposed by Deal and Grove, [6] where the oxidation time $t$ and the oxide thickness $x$ are related by:

$$t = x^2/B + x/(B/A). \qquad (1)$$

The parabolic rate $B$ is related to the oxidant diffusivity $D$ by:

$$B = 2DC_0/N, \qquad (2)$$

where $C_0$ is oxidant concentration at the surface and $N$ is the oxygen density in the oxide. The linear rate constant $B/A$ is related to the reaction rate constant at the interface $k_S$ by:

$$B/A = C_0 k_S/N. \qquad (3)$$

For Si oxidation, molecular oxygen $O_2$ has been identified as the oxidant with the activation energy of the linear rate (~2 eV) related to the barrier for $O_2$ incorporation at the advancing interface and the activation energy of the crystal-orientation-independent [7] parabolic rate (~1.2 eV) related to $O_2$ incorporation and diffusion in $SiO_2$ [8].

SiC and its oxidation have been widely studied for a range of applications, especially as a semiconductor for high-temperature, high-field, and high-power electronics, [9-12] and in oxidation-resistant coatings for high-temperature structural materials [13,14]. Similar to Si, SiC oxidation shows a linear-parabolic behavior. However, even though oxygen is again diffusing through $SiO_2$, the oxidation rate in the diffusion-limited parabolic regime is smaller and the activation energies are different. SiC also shows much stronger face-dependence than Si. For the Si-face SiC, the activation energy of the parabolic rate is ~3 eV [15-17]. For the C-face, different kinetics are observed at different temperatures. Above 1350 °C, the parabolic rate and activation energy are similar to the Si-face [17]. Below 1350 °C, smaller values for the activation energy, ~1 eV [17] and ~2 eV [15], are reported. Currently, these kinetics data, together with other experimental results, are poorly understood [18].

In this Letter, we present results from first-principles calculations to elucidate the oxidation mechanisms of SiC and contrast with the corresponding mechanisms of Si. We find that for Si-face SiC, the last layer of the oxide is very tight and inhibits the incorporation of $O_2$ at the interface, but allows atomic oxygen $O_i$, resulting in an atomic oxygen mechanism. For C-face SiC, we find that the interface inevitably contains a high density of defects, which can crack the molecule and thus facilitate a molecular mechanism. These conclusions are checked against a range of experimental data. Finally, we compare SiC oxidation to Si and propose a generic rule of how interface bonding controls the oxidation mechanism.

The most studied SiC face is the Si-face. Compared to Si, its oxidation is much slower in both the linear and the parabolic regimes and requires a higher temperature. However, the activation energy of the linear rate is only about 1.3 eV, [15,19] smaller than that of Si (~2 eV), which would suggest faster oxidation and lower temperature. The other puzzle is that the activation energy of the parabolic rate (~3 eV) is different from the diffusion barrier of $O_2$ in $SiO_2$, making it difficult to explain with the molecular mechanism. For the second puzzle, Song *et al.* proposed that the parabolic rate may be limited by the out-diffusion of gaseous oxidation products, in which case the activation energy is determined by the forward and reverse reaction rates and the diffusion barrier of the product [15]. However, since the forward and reverse reaction rates are likely to be face-dependent, the proposal does not explain the similarities between Si-face and high-temperature C-face [17].

The similarity between Si-face and high-temperature C-face [17] actually hints to the same mechanism. The authors of Ref. 17 reported that the transition of oxidation kinetics of C-face at 1350 °C coincides with an increase of lattice oxygen exchange in $SiO_2$, indicating that the oxygen diffusion mechanism changes from molecular to atomic. The authors therefore suggested atomic oxygen being the oxidant for the high-temperature C-face [20]. We note that the similarities of the Si-face and the high-temperature C-face [17] may not be coincidental, and the atomic mechanism that may account for the high temperature C-face could also be responsible for the Si-face at all temperatures.

Theoretical results in the literature also hint to an atomic mechanism. Roma *et al.* reported an activation energy of 2.8 eV for oxygen diffusion in α-quartz under extrinsic condition (i. e. the system exchanges oxygen with an $O_2$ reservoir), [21] which is close to the activation energy of the parabolic rate (~3 eV). However, to our best knowledge, no connection has been made to SiC oxidation. Furthermore, an explanation of why the more diffusive $O_2$ does not contribute to the oxidation of Si-face SiC is needed. Gavrikov *et al.* reported a barrier of 3.5 eV for $O_2$ diffusing through the last layer of $SiO_2$ on Si-face SiC, [22] which could have suggested that $O_2$ cannot be incorporated. However, they reported an even higher barrier of 4.5 eV for $O_i$ and concluded that $O_i$ is irrelevant for Si-face oxidation. Instead, they argued that the incorporation of $O_2$ is facilitated by imperfections and impurities [22]. Such an explanation, however, cannot account for the difference between the activation energy of the parabolic rate and the diffusion barrier of $O_2$.

To test whether an atomic mechanism is responsible for Si-face SiC oxidation, we performed first-principles density functional theory (DFT) calculations of the diffusion of both molecular and atomic oxygen at the interface. The calculation was done at a defect-free interface model of a (3x4) unit cell constructed by a Monte-Carlo bond-switching method bond-switching method [23]. The model consists of 8 layers of 4H-SiC with a total thickness of 16 Å and an amorphous $SiO_2$ layer with a total thickness of 10 Å. We employ the PBE version of the exchange-correlation functional, [24] PAW potentials, [25] and plane-wave basis as implemented in the VASP code [26]. We use 400 eV for the plane-wave cutoff and a shifted 2x2x1 k-point mesh. The geometrical optimizations are converged to $10^{-3}$ eV for the energy difference between two ionic steps. The diffusion barriers are calculated using the nudged elastic band method [27].

First we investigate the diffusion of an $O_2$ molecule (triplet) from the near-interface oxide to the interface (Fig. 1). In SiO$_2$, O$_2$ resides in the void of the Si-O bond network and diffuses through the passages between the voids. The energy of an interstitial O$_2$ molecule in different near-interface voids in the oxide (initial state in Fig. 1) differs by ~1 eV because of the amorphous nature of the oxide. We found a barrier of ~4 eV with a variation of ~0.3 eV for O$_2$ to penetrate the last layer. This value is close to the reported theoretical barrier of 3.5 eV for O$_2$ penetration reported in Ref. 22. The higher barrier compared to the case of Si oxidation (~2 eV [8]) is consistent with the shorter length of Si-C bonds compared to the Si-Si bonds in Si, which causes a higher density of Si-O bonds when the surface Si atoms are satisfied, thus resulting in smaller voids and narrower passages. At typical oxidation temperature, the high barrier blocks the diffusion of O$_2$ and makes it an ineffective oxidant.

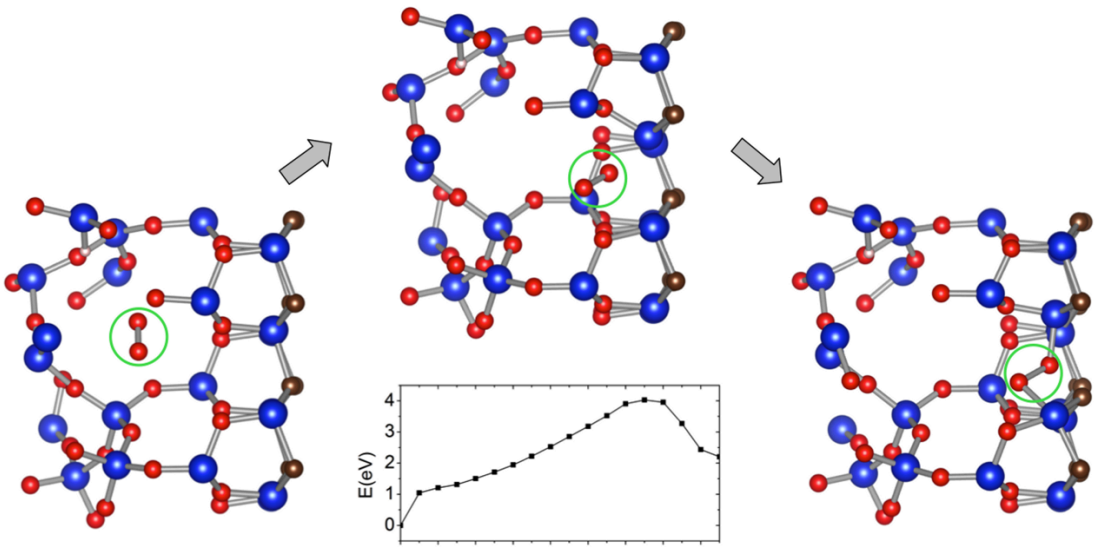

Fig. 1. The initial, transition (top of the energy barrier), and final states of an O$_2$ penetrating the last layer of SiO$_2$. Si in blue, C in brown, O in read, and H (for passivating the dangling bonds in the model) in white. The green circle highlights the diffusing oxidant. The energy profile along the reaction path is also shown.

For atomic oxygen O$_i$, the situation is different. In SiO$_2$, an O$_i$ takes the form of Si-O-O-Si peroxide linkage (bridge interstitial in a Si-O bond) and diffuses by hopping from one bond to another [28]. At the interface, an O$_i$ can hop onto the first layer of SiC and take the form of bridge interstitial in a Si-C bond. Such diffusion modes do not require passages between the voids and is not hampered by the higher density of the Si-O bonds. For an O$_i$ to reach the surface of SiC (Fig. 2), we obtained a barrier of 1.2 eV with a variation of 0.1 eV, caused by the amorphous nature of the oxide. This value is lower than the energy barrier $E_{Diff}$ for atomic oxygen diffusion in SiO$_2$ away from the interface, which is 1.7 eV (+/− 0.1 eV) based on our calculation. As a result, atomic oxygen is not blocked, and can arrive at the surface and oxidize the SiC.

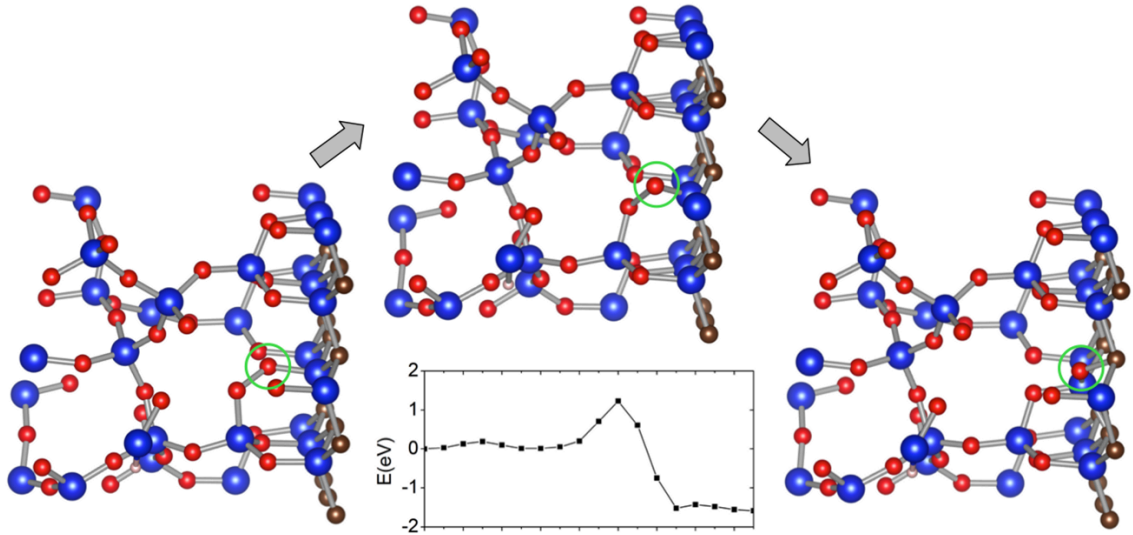

Fig. 2. The initial, transition, and final states of an $O_i$ arriving at the surface of SiC. The energy profile along the reaction path is also shown.

During thermal oxidation, atomic oxygen can be generated at the oxide surface through the dissociation of $O_2$ ($O_2 \rightarrow 2O_i$). A schematic description of the atomic oxygen mechanism is shown in Fig. 3. Using results from first-principles calculations, we check the proposed mechanism against the experimental data. First we look at the activation energies. According to Deal and Grove, the parabolic rate constant $B$ equals $DC_0/N$, (the factor 2 difference from Eq. (2) is because the oxidant now contains one oxygen atom instead of two), where $D$ is proportional to $\exp(-E_{Diff}/kT)$. Assuming the reaction $O_2 \leftrightarrows 2O_i$ is at equilibrium near the surface, $C_0$ is determined by the $O_i$ formation energy $E_{Form}$ through $C_0 = N\exp(-E_{Form}/kT)$, where $N$ is the density of the possible $O_i$ sites that happens to equal the oxygen density. The activation energy of $B$ is therefore $E_{Diff} + E_{Form}$. We calculated the formation energy $E_{Form}$ of an $O_i$ near the surface and obtained values between 1.3 to 1.5 eV. Combining this value with $E_{Diff}$ = 1.7 eV, we obtained an activation energy of 3.1 eV, consistent with the experimental value of ~3 eV.

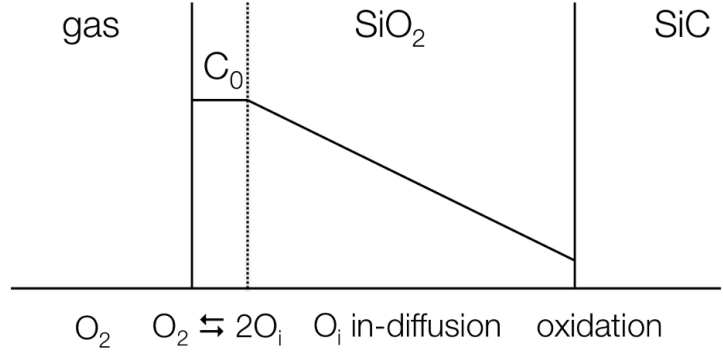

Fig. 3. Atomic oxygen concentration profile of the atomic oxygen mechanism.

Now let us discuss the activation energy of the linear rate. According to Eq. (3), the linear rate constant $B/A$ equals $C_0 k_S/(4N)$, (the factor 4 difference from Eq. (3) is because the oxidant now contains one oxygen atom instead of two). In the case of Si oxidation, $k_S$ is temperature-dependent and its activation energy equals the energy barrier for $O_2$ penetrating the last layer of oxide. This originates from the fact that the energy barrier for $O_2$ to penetrate the last layer of oxide (~2 eV) is larger than the energy barrier for its diffusion in bulk $SiO_2$ (1.2 eV) [8]. This higher barrier at interface can result in an accumulation of $O_2$ and limit the reaction rate $k_S$. However, in the case of the oxidation of Si-face SiC, the energy barrier for $O_i$ to penetrate the last oxide layer (~1.2 eV) is smaller than the barrier for it to diffuse in bulk $SiO_2$ (~1.7 eV). As a result, the $O_i$'s never accumulate at the interface because of the energy barrier. Thus, on the Si-face of SiC, $k_S$ is not limited by the energy barrier of $O_i$ penetration at the last layer. Therefore we assume that $k_S$ is not strongly dependent on temperature, whereby $B/A$ is proportional to $C_0$, and its activation energy should equal $E_{Form} = 1.4$ eV. This prediction agrees well with the experimental value of ~1.3 eV, [15,19] suggesting the validity of the assumption and the atomic mechanism.

The atomic oxygen mechanism not only predicts the activation energies, but also predicts the rate constant $B$. Recall that $B$ equals $DC_0/N$, and $D$ is related to $E_{Diff}$ by $D = 1/6 l^2 f \exp(-E_{Diff}/kT)$, where $l$ is the distance of each hop and $f$ is the attempt frequency [29]. Using $l = 2.7$ Å (the average distance between nearest-neighbor oxygen atoms in the $SiO_2$ part of our model), $f = 10^{13}$ s$^{-1}$, $E_{Form} = 1.4$ eV, and $E_{Diff} = 1.7$ eV, we obtain $B \sim 1.3 \times 10^{-14}$ cm$^{-2}$/s at 1150 °C. This value is within the reported experimental values, which range from $1.1 \times 10^{-15}$ to $3.0 \times 10^{-14}$ cm$^{-2}$/s (deduced from Ref. 15, Ref. 16, and Ref. 30).

The atomic oxygen mechanism also explains the pressure dependence. Assuming the interface reaction rate constant $k_S$ is independent of the pressure, the molecular mechanism predicts $B/A$ being proportional to the pressure of the $O_2$ gas, while the atomic mechanism predicts *B/A* being proportional to the square root of the $O_2$ pressure. Experimentally, sub-linear dependence of *B/A* is observed, which fits to the power of 0.4, [31] close to the predicted value of 0.5.

Besides the kinetic data, the atomic mechanism also explains the double oxidation experiments that are not well understood so far [32,33]. In those experiments, $SiO_2$ is grown on SiC first by $^{16}O_2$ gas and then by $^{18}O_2$ gas. On the Si-face, the new oxide grown in the second oxidation has only a small percentage of $^{18}O$, which is difficult to explain by the molecular mechanism, but is a natural consequence if atomic oxygen is the oxidant: $O_i$ diffuses in $SiO_2$ through a kick-out mechanism, in which it exchanges with a lattice oxygen. Therefore, during the second oxidation, as a flux of $O_i$'s move towards the interface, the percentage of $O_i$ being $^{18}O$ rapidly decreases. When the flux reaches the interface, most $O_i$'s are $^{16}O$ that are originally part of the lattice but have been kicked-out. As a result, oxygen atoms in the newly grown oxide are predominantly $^{16}O$, consistent with observations.

The fact that the oxygen atoms in the new oxide are originally lattice oxygens can be seen from the triple oxidation experiment, which has a third oxidation with $^{16}O_2$ after the second oxidation with $^{18}O_2$. After the third oxidation, the newest oxide contains a fairly high concentration of $^{18}O$, [34] which can only come from the oxide near the interface.

We now discuss the oxidation of C-face SiC. To explore oxygen diffusion at the C-face, we need to know about the structure of the stable interface. It may seem natural to assume that the C-face SiC/$SiO_2$ interface terminates at the C-layer, with SiC connected to $SiO_2$ by C-O bonds, in analogy to the Si-face interface model shown in Fig. 1 and 2. However, such a structure is not stable at high temperature, as carbon atoms in contact with oxygen are emitted as CO or $CO_2$. Thus the interface terminates at the Si layer instead. At this layer, each Si atom is bonded to the SiC substrate with only one Si-C bond, leaving 3-bonds to be satisfied. Meeting such requirement with Si-O bonds would result in an impossibly high oxide density. Instead, the C-face must be oxygen deficient and have a high density of Si-Si bonds (oxygen vacancies). This conclusion is consistent with the SIMS profile of as-oxidized C-face SiC, which qualitatively shows a Si excess at the interface [35]. The Si-Si bonds crack the $O_2$ molecule and result in the growth of $SiO_2$. We calculated one of such reactions (Fig. 4) and obtained an energy barrier of only 0.9 eV, which confirms that a defective interface layer enables molecular oxygen to be an effective oxidant. [36] We also explored the incorporation of atomic oxygen into the C-face SiC and found a small energy barrier of 0.35 eV. Therefore atomic oxygen can also oxidize C-face SiC. However, since $O_2$ is more abundant at the interface at normal oxidation temperature (below 1350 °C), [17, 20] the main contribution to the oxidation kinetics of the C-face comes from the molecular mechanism.

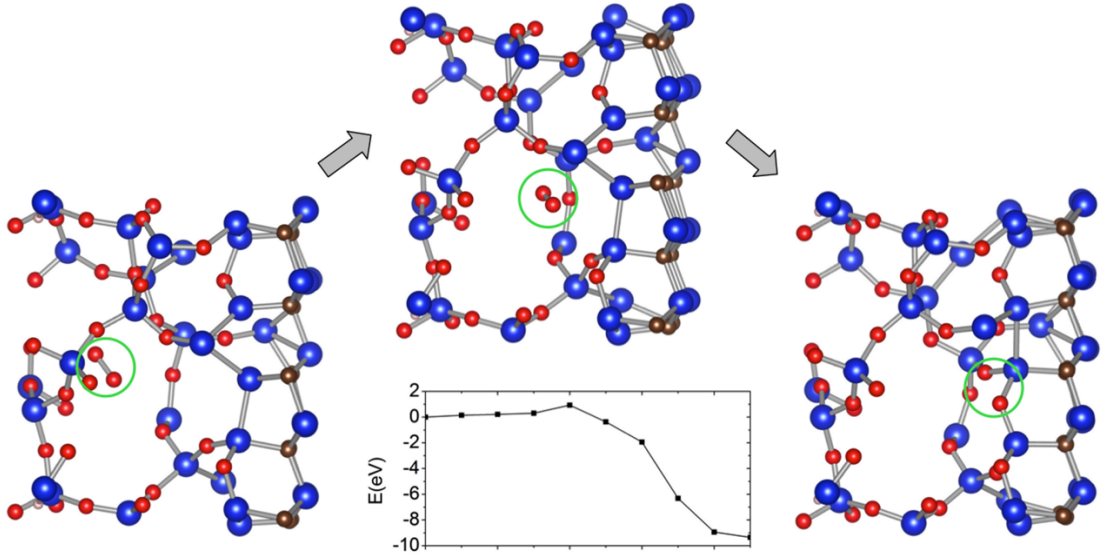

Fig. 4. The initial, transition, and final states of an $O_2$ reacting with a defective layer at the C-face $SiC/SiO_2$ interface. The energy profile along the reaction path is also shown.

The kinetic data on the C-face SiC below 1350 °C are not well converged. For the parabolic rate, activation energies of ~ 1 eV [17] and ~ 2 eV [15] are reported; for the linear rate, both a single activation energy ~1.3 eV across the whole temperature range [15] and two activation energies of 0.75 and 1.76 eV in two temperature ranges [19] are reported. We therefore refrain from connecting them to processes at the atomistic scale. Nevertheless, compared to Si-face SiC, the double oxidation experiments at C-face SiC show much higher enrichment of $^{18}O$ at the second oxidation, [32-34] consistent with molecular oxygen mechanism being the major contributor. In addition, the linear rate on C-face increases linearly with the pressure, [35] also supporting the molecular mechanism.

Now let us compare the oxidation of both Si- and C-face of SiC with that of Si. Since SiC has the same native oxide as Si, under typical oxidation conditions, there should be plenty of $O_2$ molecules in the near-interface region of the oxide. In the case of Si, the interface is largely perfect with a low defect density comparing to the density of interfacial sites. The oxide layer near the interface is not too dense and $O_2$ can overcome a barrier. For Si-face SiC, the smaller lattice constant dictates a higher density of interface bonds. However, it is still possible to form a largely perfect interface. Thus there exists a denser interfacial oxide layer that filters out $O_2$ but lets through atomic oxygen. In the case of C-face SiC, the requirement of bonds at the interface is even higher, making it impossible to have a near-perfect interface and resulting in a non-negligible interface defect density comparing to the density of interfacial sites. These interface defects then enable molecular oxygen to be an effective oxidant again. The three cases represent three types of oxidation depending on the bonding requirement at the interface: direct molecular oxygen mechanism when the required density of bonds is low, atomic oxygen mechanism when the required density of bonds is high, and defect-assisted molecular mechanism when the required density of bonds is very high.

Identifying the origin of the atomic-scale oxidation mechanism provides guidance for choosing an effective oxidation method. Materials that have relatively high density of interface bonds with its native oxide (but not too high to make the interface defective) are not likely to be oxidized easily by $O_2$. We showed that a dense oxide layer that is effective against the diffusion of $O_2$ does not necessarily stop atomic oxygen. In such cases, using atomic oxygen as the oxidant would be beneficial.

In summary, we show that in contrast to Si, the Si-face SiC oxidation is dominated by an atomic oxygen mechanism. Such effect originates from the fact that the required density of bonds at the Si-face SiC/$SiO_2$ interface is higher than that at the Si/$SiO_2$ interface. For the C-face, we found that as the bonding requirement is too high to satisfy, the interface is intrinsically defective, leading to an oxidation mechanism again dominated by molecular oxygen. These three cases represent three regimes where the bonding requirements control the oxidation mechanisms, which should be considered when either improving oxide and interface, or facilitating oxidation, or improving oxidation resistance is of interest.

**ACKNOWLEDGMENTS**

The work was supported by NSF under Grant # DMR-0907385, by DOE Basic Energy Sciences, and by the McMinn Endowment at Vanderbilt University. Computational support was provided by the NSF XSEDE under Grant # TG-DMR100022. We thank Prof. L. C. Feldman for helpful discussions.

REFERENCES
[1] L. C. Feldman, E. P. Gusev, and E. Garfunkel, in *Fundamental Aspects of Ultrathin Dielectrics on Si-Based Devices*, edited by E. Garfunkel. E. Gusev, and A. Vul' (Kluwer, Dordrecht, 1998), p. 1.
[2] K. S. Novoselov, A. K. Geim, S. V. Morozov, D. Jiang, Y. Zhang, S. V. Dubonos, I. V. Grigorieva, and A. A. Firsov, *Science* 22, 666 (2004)
[3] X. Jiang, T. Herricks, and Y. Xia, Nano Lett. 2, 1333 (2002)

[4] M. N. Rahaman, *Ceramic Processing and Sintering* (CRC Press, 2003).

[5] G. W. Meetham, M. H. Van de Voorde, *Materials For High Temperature Engineering Applications* (Springer, New York, 2000).

[6] B. E. Deal and A. S. Grove, *J. Appl. Phys.* **36**, 3770 (1965).

[7] C. Hollauer, *PhD Thesis*. (Vienna University of Technology, Vienna, 2007).

[8] L. Tsetseris and S. T. Pantelides, *Phys. Rev. Lett.* **97**, 116101 (2006).

[9] J. Cooper and A. Agarwal, *Proc. IEEE* **90**, 956 (2002).

[10] M. Di Ventra and S. T. Pantelides, *Phys. Rev. Lett.* **83**, 1624 (1999).

[11] S. Wang, M. Di Ventra, S. G. Kim, and S. T. Pantelides, *Phys. Rev. Lett.* **86**, 5946 (2001).

[12] S. Wang, S. Dhar, S-R. Wang, A.C. Ahyi, A. Franceschetti, J.R. Williams, L.C. Feldman and S.T. Pantelides, *Phys. Rev. Lett.* **98**, 026101 (2007).

[13] F. Smeacetto, M. Salvo, and M. Ferraris, *Carbon* **40**, 583 (2002)

[14] J.-F. Huang, X.-R. Zeng, H.-J. Li, X.-B. Xiong, and Y.-W. Fu, *Carbon* **42**, 1517 (2004)

[15] Y. Song, S. Dhar, L. C. Feldman, G. Chung, and J. R. Williams, *J. Appl. Phys.* **95**, 4953 (2004).

[16] B. K. Daas, M. M. Islam, I. A. Chowdhury, F. Zhao, T. S. Sudarshan, M. V. S. Chandashekhar, *IEEE Trans. Electron Devices* **58**, 115, (2011).

[17] Z. Zheng, R. E. Tressler, and K. E. Spear, *J. Electrochem. Soc.* **137**, 854 (1990).

[18] I. Vickridge, J. Ganem, Y. Hoshino, and I. Trimaille, *J. Phys. D.* **40**, 6254 (2007).

[19] T. Yamamoto, Y. Hijikata, H. Yaguchi, and S. Yoshida, *Jpn. J. Appl. Phys.* **47**, 7803 (2008).

[20] Z. Zheng, R. E. Tressler, and K. E. Spear, *J. Electrochem. Soc.* **137**, 2812 (1990).

[21] G. Roma, Y. Limoge, and S. Baroni, *Phys. Rev. Lett.* **86**, 4564 (2001).

[22] A. Gavrikov, A. Knizhnik, A. Safonov, A. Scherbinin, A. Bagatur'yants, B. Potapkin, A. Chatterjee, and K. Matocha, *J. Appl. Phys.* **104**, 093508 (2008).

[23] Y. Tu and J. Tersoff, *Phys. Rev. Lett.* **84**, 4393 (2000).

[24] J. Perdew, K. Burke, and M. Ernzerhof, *Phys. Rev. Lett.* **77**, 3865 (1996).

[25] G. Kresse and D. Joubert, *Phys. Rev. B* **59**, 1758 (1999).

[26] G. Kresse and J. Furthmuller, *Phys. Rev. B* **54**, 11169 (1996).

[27] G. Henkelman and H. Jonsson, *J. Chem. Phys.* **113**, 9978 (2000).

[28] D. R. Hamann, *Phys. Rev. Lett.* **81**, 3447 (1998).

[29] P. E Blöchl, Enrico Smargiassi, R. Car, D. B. Laks, W. Andreoni, and S. T. Pantelides , *Phys. Rev. Lett.* **70**, 2435 (1993).

[30] X.-A. Fu, K. Okino, and M. Mehregany, *Appl. Phys. Lett.* **98**, 042109 (2011)

[31] E. A. Ray, J. Rozen, S. Dhar, L. C. Feldman, and J. R. Williams, *J. Appl. Phys.* **103**, 023522 (2008)

[32] C. Radtke, I. J. R. Baumvol, B. C. Ferrera, and F. C. Stedile, *Appl. Phys. Lett.* **85**, 3402 (2004)

[33] I. C. Vickridge, J. J. Ganam, I. Trimaille, J.-L. Cantin, *Nucl. Instr. Methods B*, **232**, 272 (2005).

[34] I. Trimaille, J. J. Ganem, I. C. Vickridge, S. Rigo, G. Battistig, E.Szilagyi, I. J. R. Baumvol, C. Radtke, F. C. Stedile, *Nucl. Instr. Methods B*, **219-220**, 914 (2005).

[35] S. Dhar, S. T. Pantelides, J. R. Williams, and L. C. Feldman, in *Defects in Microelectronic Materials and Devices,* edited by D. M. Fleetwood, S. T. Pantelides, R. D. Schrimpf (CRC Press, Boca Raton, FL, 2009), p. 575.

[36] As the Si-layer gets oxidized, it relaxes into $SiO_2$, decreasing the volume density of Si atoms at the interface. This makes the next C-layer under-coordinated and suspect to further removal by oxygen. After the removal of C-layer, the interface again terminates at the next Si-layer.